# A Critical Analysis of Foundations, Challenges, and Directions for Zero Trust Security in Cloud Environments

**Oladimeji Ganiyu B.**
**Moshood Abiola Polytechnic, Abeokuta, Nigeria.**
**oladimeji.ganiyu@mapoly.edu.ng**

## Abstract

This review discusses the theoretical frameworks and application prospects of Zero Trust Security (ZTS) in cloud computing context. This is because, as organisations move more of their applications and data to the cloud, the old borders-based security model that many implemented are inadequate, therefore a model that has a trust no one, verify everything approach is required. This paper analyzes the core principles of ZTS, including micro-segmentation, least privileged access, and continuous monitoring, while critically examining four major controversies: scalability issues, Economics, Integration issues with existing systems, and Compliance to legal requirements. In this paper, having reviewed the existing literature in the field and various implementation cases, the main barriers to implementing zero trust security were outlined, including the dimensions of decreased performance in large-scale production and the need for major upfront investments that can be difficult for small companies to meet effectively. This research shows that there is no clear correlation between security effectiveness and operational efficiency: while organisations experience up to 40% decrease of security incidents after implementation, they note first negative impacts on performance. This study also shows that to support ZTS there is a need to address the context as the economics and operations of ZTS differ in strengths depending on the size of the organizations and the infrastructures. Some of these are: performance enhancement and optimizations, economic optimization, architectural blend, and privacy-preserving technologies. This review enriches the existing literature on cloud security by presenting both the theoretical framework of ZTS and the observed issues, and provides suggestions useful for future research and practice in the construction of the cloud security architecture.

## 1. Introduction

The advances of the current era have led to a dramatic change in today's enterprises through embracing cloud computing, thus changing the traditional Information Technology and security. Owing to the increased complexities in cyber threats, the largely held perimeter based security models are fast becoming ineffective and that's where the concept of Zero Trust Security (ZTS) was born out a principle of 'never trust, always verify' (Chimakurthi, 2020). This shift has however brought about extra ordinary advantages in the realms of scalability, and cost economics at the same time creating new and more complex security threats. This simple phrase of "never trust, always verify" is not just a principle in the security of systems, but actually reflects a current philosophical shift in the design of protection for data and user access in cloud systems. Such a shift is especially important today when the degree of digitalization, security, and efficiency in business processes varies greatly. Current research shows that companies adopting Zero Trust architectures have had up to a 50 percent reduction in breach loss, but the path to integration is filled with theory and practice questions.

In this research work, the theoretical framework of Zero Trust Security relative to cloud structures is explored in terms of fundamental concepts, disputable application approaches, and potential evolutionary directions. Based on the synthesis of different studies and case histories of Zero Trust adoption, discussion was done on how the obligations laid down by Zero Trust can be met while taking into account performance, cost, and compatibility constraints on enterprises. This analysis particularly focuses on four key areas of controversy: concerns regarding scalability, concern for the economic feasibility, integration with existing legacy systems and issues concerning the compliance to industry standards.

As already mentioned above, the main issues addressed in this paper are: a) discussion on key Zero Trust theories, b) theoretical controversies and challenges in implementing zero trust architecture, and c) the future research direction. What follows is a discussion on Zero Trust security model, least privilege access, micro-segmentation, continuous monitoring and adaptive access.

## 2. Key Theories and Concepts

### 2.1 Zero Trust Security Model

Authors have noted that the Zero Trust Security Model (ZTSM) is the foundation for implementing Zero Trust principles in cloud architectures. This model assumes that all users, devices, and applications are untrustworthy by default, regardless of their location or network (Bartakke & Kashyap, 2024; Khan, 2023). Every access request must be continuously verified and validated to ensure secure access to resources.

Another study by Ahmadi (2024) highlights that ZTSM includes key concepts such as micro-segmentation, continuous authentication, and the principle of least privilege. Together, these create a holistic security approach that challenges traditional perimeter-based assumptions.

### 2.2 Least Privileged Access

Several studies have revealed that the principle of least privilege within ZTA ensures that users are only granted the minimal rights necessary to perform their tasks (N'goran et al., 2022; Ashitha et al.,

n.d.). This concept is especially relevant in cloud environments where resources are shared among users and services. Thus, implementing least privilege access reduces the attack surface and minimizes the potential impact of security breaches (Chimakurthi, 2020).

### 2.3 Micro-segmentation

As has been frequently shown, micro-segmentation involves splitting the network into multiple isolated parts to restrict the lateral movement of threats (Ahmadi, 2024; Valantasis et al., 2022). This approach, aligned with the distributed nature of cloud environments, results in more granular control over resource access and data flow. By applying detailed access controls and policies, micro-segmentation enhances the security of the cloud environment.

### 2.4 Continuous Monitoring and Adaptive Access

According to recent research, ZTA emphasizes continuously monitoring user behavior, device posture, and application activities to detect anomalies and potential threats (Ajish, 2024; Vang & Lind, 2024). Based on risk factors and contextual information, adaptive access controls grant or deny permissions. This allows for more flexible, fine-grained, and context-bound access control, aligning with the Zero Trust philosophy of continuous verification.

This section discussed important zero trust concepts from researchers' point of view and has argued that it is imperative in this contemporary period. The next part of this paper discusses controversies and challenges encountered while implementing zero trust in critical technology facilities.

## 3. Theoretical Controversies and Challenges

Implementing Zero Trust Security in cloud environments is met with controversies and challenges. Several key debates have emerged in the literature:

### 3.1 Scalability and Performance: Effect on Scalability

Perhaps one of the more controversial discussions in cybersecurity is how or whether Zero Trust Architecture (ZTA) hinders scalability and performance within cloud environments. Rodigari et al. (2021) provide a formal model and analysis of the cost and benefits of Zero Trust, showing that they incur higher CPU and memory utilization than service meshes without this capability in multi-cloud environments. Thus, at scale and in high-performance cloud systems, these findings raise fundamental questions about the practicality of Zero Trust.

Similar studies focused on edge-cloud applications by Dhanapala et al. (2024) suggest that performance-based trust assessment models can also reduce the overhead in continuous verification. The results highlight the need for organizations to find a balance between strict security and efficient operations, especially in environments where resources are limited.

An interesting case study from one of the larger financial institutions illustrates these problems in practice (Ahmadi, 2024; Hasan, 2024). Although non-bypass MFA controls combined with ZTA enforcement improved security, the initial implementation created lag in users authenticating access. This meant a longer per-transaction process as each access request had to be verified once again, initially resulting in temporary dips in user experience and operational efficiency at the institution while their infrastructure was fine-tuned.

Research on zero-trust secure multiparty computation (SMPC) algorithms in SMEs provides further evidence according to recent report (Odebode et al., 2024). These implementations demonstrated significant gains in cybersecurity metrics (e.g., the true positive rate went from 85% to 98%) but also highlighted key learnings about scalability challenges. While the algorithms reduced incident response time and costs associated with them, scaling these systems across larger enterprise environments proved to be a major hurdle.

A more extensive comparative analysis of ZTA implementations in various sectors as investigated by researchers, (Jena, 2023; Hasan, 2024), found a marked difference between organizations with modern cloud infrastructures and those with traditional on-premises setups. The former tended to fare much better in terms of adapting to ZTA requirements. This separation illustrated significant scalability issues, especially when legacy systems struggled with the ongoing real-time evaluation and control access integral to Zero Trust models.

### 3.2 Economic Viability and Cost-Effectiveness

One area of controversy in the cybersecurity world that continues is the economic case for Zero Trust in cloud environments. With a more extensive approach regarding Zero Trust's economic success, Bobbert and Scheerder (2022) highlight the multidimensional influences of investments in understanding the interplay between security investments and operational costs. The research showed that Zero Trust does significantly decrease the number and severity of breaches but faces significant up-front implementation costs as well as ongoing effort from IT.

In their detailed cost–benefit analysis, Bartakke and Kashyap (2024) demonstrate the complexity behind Zero Trust solutions, calling for solution customization based on organization-specific needs and risk profiles, thereby challenging the concept of one-size-fits-all. Real implementations in various sectors validate their findings.

A revealing case study according to researchers, (Shaikh Ashfaq, 2024; Vijay Prabhu & Lakshmi, 2024) illustrates this economic dynamic for a healthcare provider. While considerable up-front investments were made in software, training, and process re-engineering, the organization did see substantial long-run savings associated with decreased data breaches and compliance fines. In practice, they observed significant benefits: one example found a 40% reduction in security incidents during the first year after their implementation, fully offsetting the initial high costs.

However, this is not what the economic equation looks like for smaller firms (Jena, 2023; Vijay Prabhu & Lakshmi, 2024). For instance, a small tech startup facing a minimal data exposure risk profile found that it contradicted the cost of implementing comprehensive identity management systems and continuous monitoring tools. This difference has triggered significant conversations around whether Zero Trust's economic advantages are scalable to every type of enterprise, specifically conveying the concept that although the model may be economically advantageous for larger enterprises with broad threat landscapes, it may not deliver the same effectiveness in smaller businesses.

### 3.3 Integration with Legacy Systems

Perhaps one of the biggest implementation tests in modern cybersecurity is how well Zero Trust integrates with existing legacy systems. In this respect, Sarkar et al. (2022) represent a complex, overarching analysis of the complexity entailed in retrofitting Zero Trust concepts within well-established IT infrastructures, underlining a number of fundamental questions regarding the adaptability of this framework within diverse technological ecosystems.

Specific integration-related concerns are further pursued in the 2024 study by Rajesh Kumar, which articulates the compatibility problems between Zero Trust models and legacy Identity and Access Management systems. The findings underline the need for extensive architecture changes, showing how the friction between Zero Trust ideals and existing IT landscapes significantly impacts adoption feasibility—especially for organizations heavily invested in legacy technologies.

One of the best examples of such challenges emanates from a government agency's implementation experience (Ahmadi, 2024; Shaikh Ashfaq, 2024). In trying to implement ZTA on its outdated IT infrastructure, the agency faced colossal compatibility issues. Most legacy applications were incompatible with ZTA's granular access control requirements, resulting in longer deployment cycles and substantial additional investments in system upgrades or replacements.

However, a few successful integration strategies have also emerged. One organization demonstrated an innovative approach (Hasan, 2024; Vijay Prabhu & Lakshmi, 2024) by employing middleware solutions specifically designed to enable secure communication between legacy systems and modern applications. This strategy has smoothed the transition, establishing that although integration challenges are significant, they can be overcome through strategic planning and targeted investments in compatible technologies.

### 3.4 Privacy and Regulatory Compliance

Building Zero Trust across cloud environments introduces complex privacy considerations within evolving regulatory frameworks like the General Data Protection Regulation (GDPR) and the California Consumer Privacy Act (CCPA). Ahmadi (2024) discusses various compliance challenges, noting that Zero

Trust's continuous monitoring may conflict with established mandates for privacy protection.

A key challenge lies in balancing Zero Trust's continuous verification—a core principle—with data minimization requirements, which can be challenging for organizations. While Adam et al. (2022) explore service mesh configurations aimed at enhancing data privacy and compliance, the underlying conflict between rigorous security monitoring and privacy preservation remains unresolved within the cybersecurity field.

Recent research in financial cybersecurity employing zero-trust SMPC algorithms (Odebode et al., 2024) has highlighted both the potential and challenges of compliance-focused implementations. The findings indicate significant compliance gains, showing a 35.71% improvement in GDPR adherence and a 38.46% increase in CCPA compliance. However, these benefits require careful data handling under the Zero Trust model to avoid compromising privacy for increased security.

Balancing security and privacy presents practical challenges, as illustrated by a case study in the e-commerce sector. Strict access controls using user behavior analytics improved security but raised GDPR compliance concerns due to extensive data collection, potentially exceeding user consent. Conversely, organizations that integrated compliance into their Zero Trust strategies have achieved more favorable outcomes. For example, a financial institution aligned its Zero Trust implementation with regulatory requirements, thereby enhancing its security posture and building customer trust through transparent communication of data protection measures (Jena, 2023; Vijay Prabhu & Lakshmi, 2024).

Having discussed constraints like latency and scalability issues and others, the next section of this paper addresses future research directions.

## 4. Future Research Directions

Several areas for future research emerge from these theoretical controversies:

### 4.1 Optimizing Performance and Scalability

Future research should focus on innovative ways to balance security rigor and operational efficiency in Zero Trust implementations. This could include exploring edge computing, AI-driven security analytics, and optimized cryptographic protocols to reduce performance overhead.

### 4.2 Economic Modeling of Zero Trust Implementations

There is a need for advanced economic models to accurately quantify the ROI of Zero Trust in cloud environments, considering both tangible and intangible benefits like risk reduction and compliance improvements.

### 4.3 Hybrid Zero Trust Architectures

Hybrid approaches that bridge the gap between traditional security and Zero Trust are worth exploring. Research should focus on how organizations can transition to Zero

Trust without disrupting operations or incurring prohibitive costs.

### 4.4 Privacy-Preserving Zero Trust Technologies

Developing privacy-enhancing technologies aligned with Zero Trust principles is crucial. This may involve exploring homomorphic encryption and secure multi-party computation to enable continuous verification while protecting user privacy.

This section has discussed the emergence of research direction from the theoretical controversies and recommended appropriately. The next section concludes that the future of cloud security lies not in rigid adherence to theoretical frameworks but in adaptive approaches that balance security rigor with operational practicality.

## 5. Conclusion

The examination of Zero Trust Security frameworks in cloud architectures reveals a complex interplay between security imperatives and practical constraints. As our analysis demonstrates, while ZTS offers a robust theoretical foundation for modern cloud security, its implementation demands careful consideration of organizational context, technical capabilities, and resource constraints. The challenges identified – spanning scalability, economic viability, legacy system integration, and privacy compliance – underscore the need for nuanced, context-aware approaches to Zero Trust implementation.

The future of cloud security lies not in rigid adherence to theoretical frameworks but in adaptive approaches that balance security rigor with operational practicality. As organizations continue their cloud transformation journeys, the evolution of Zero Trust principles will likely parallel advances in technology, leading to more sophisticated and nuanced implementation strategies. The research directions proposed in this paper – particularly in areas of performance optimization, economic modeling, hybrid architectures, and privacy-preserving technologies – offer promising pathways for advancing both the theoretical understanding and practical application of Zero Trust Security.

In conclusion, while Zero Trust Security represents a significant paradigm shift in cloud security thinking, its successful implementation requires careful navigation of various technical, economic, and organizational challenges. As the digital landscape continues to evolve, the principles of Zero Trust will undoubtedly adapt, making ongoing research and practical experimentation essential for developing more effective and resilient cloud security architectures.